%% file: main.tex
\documentclass[sigconf,authorversion]{acmart}

\AtBeginDocument{%
  \providecommand\BibTeX{{%
    \normalfont B\kern-0.5em{\scshape i\kern-0.25em b}\kern-0.8em\TeX}}}

\usepackage{amsmath}
\usepackage{multicol}
\usepackage{multirow}

\usepackage{siunitx}
\usepackage{tabularx}
\usepackage[shortlabels]{enumitem}

\usepackage{lscape}

\author{Chen Liang}
\email{lliangchenc@163.com}
\email{liang-c19@mails.tsinghua.edu.cn}
\orcid{0000-0003-0579-2716}
\affiliation{
  \institution{Pervasive Computing Group, Tsinghua University}
  \country{Beijing, China}
}

\begin{document}

\title[]{Towards Ubiquitous Intelligent Hand Interaction}



\renewcommand{\shortauthors}{Liang et al.}

\begin{abstract}
The development of ubiquitous computing and sensing devices has brought about novel interaction scenarios such as mixed reality and IoT (e.g., smart home), which pose new demands for the next generation of natural user interfaces (NUI). Human hand, benefit for the large degree-of-freedom, serves as a medium through which people interact with the external world in their daily lives, thus also being regarded as the main entry of NUI. Unfortunately, current hand tracking system is largely confined on first perspective vision-based solutions, which suffer from optical artifacts and are not practical in ubiquitous environments. In my thesis, I rethink this problem by analyzing the underlying logic in terms of sensor, behavior, and semantics, constituting a research framework for achieving ubiquitous intelligent hand interaction. Then I summarize my previous research topics and illustrated the future research directions based on my research framework.
\end{abstract}

\begin{teaserfigure}
    \centering
    \includegraphics[width=\textwidth]{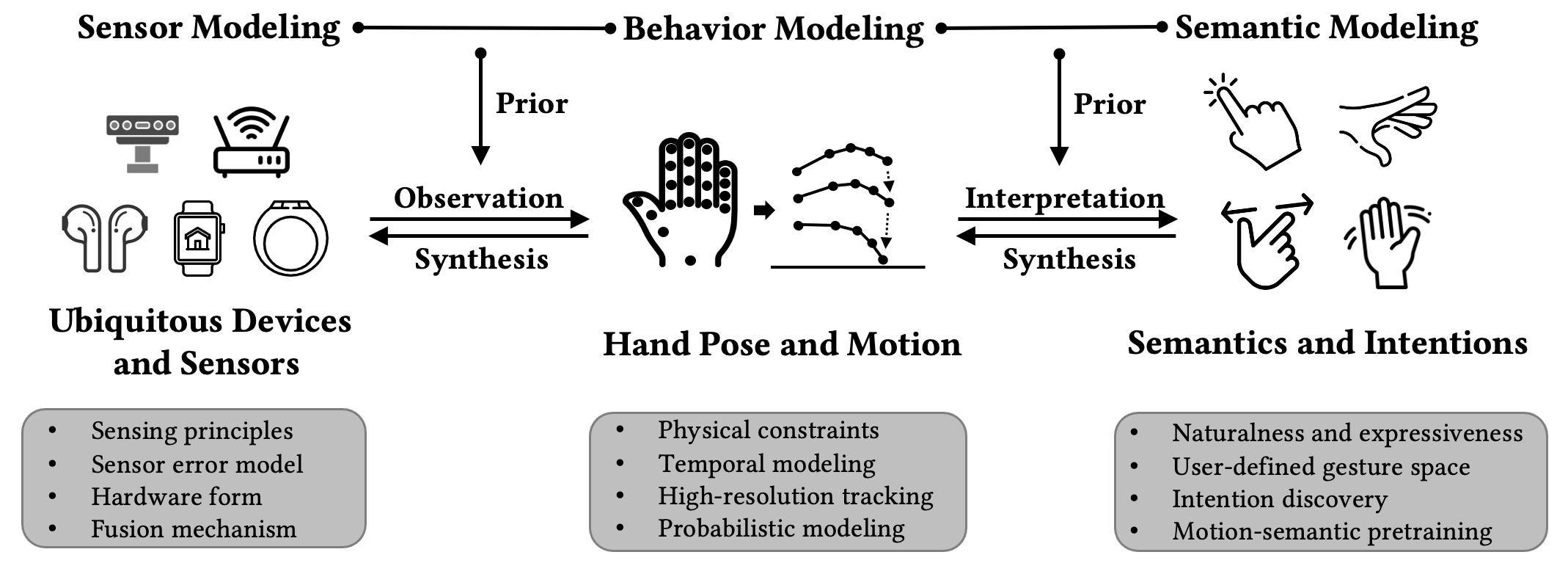}
    \caption{The research framework towards ubiquitous intelligent hand interaction.}
    \label{fig:teaser}
\end{teaserfigure}






\maketitle
\input{1_introduction}
\input{2_previous_work}
\input{3_future_work}
\input{4_conclusion}



\balance{}

\bibliographystyle{ACM-Reference-Format}
\bibliography{sample-base}



\end{document}

%% file: 1_introduction.tex
\section{Introduction}

Hand behavior sensing and modeling is an extensively researched task in computer vision and ubiquitous computing. 
Although hand tracking has been incorporated as a key input technique for mixed reality platforms, such as Microsoft Hololens \cite{hololens2} and Oculus Quest \cite{Abdlkarim2022.02.18.481001}, supporting highly usable hand interaction in the wild is still a challenging problem, especially reflected in two aspects. 

First, the current camera-based hand tracking suffered from the inherent weakness of computer vision, specifically, for environmental issues (e.g., low light or confusing background \cite{deng2021hand}), structural issues (e.g., obstruction or peripheral viewpoint \cite{cheng2022efficient}), and resolution issues (e.g., fast or subtle gestures \cite{10.1145/3332165.3347947}). For example, the very basic and important input event of transient finger-to-surface touch can hardly be distinguished from a false positive (e.g., pretending to touch) by cameras because a touch event typically occurs within 20 milliseconds and with a submillimeter spatial resolution (e.g., touch v.s. slight hovering). Similarly, there still remain a number of unsolved problems that are fundamental to facilitate intelligent hand interaction, such as free-form touch (e.g., recognizing the touch point, the touch stroke, and the hand/finger entity) on unmodified surface, transient and subtle hand gesture recognition, and Midas problem for mid-air hand gesture. 

Second, vision-based hand tracking relies on a first-perspective always-on camera and peripheral hardware with substantial physical and computational load, which is not practical for ubiquitous scenarios such as outdoor and IoT scenarios. When such a camera-based hand tracking system is absent, little can be done to support natural and expressive hand input.

Aiming to bridge the abovementioned gaps, I proposed a research framework, as shown in Figure \ref{fig:teaser}, to outline the underlying logic behind the ultimate goal to enable highly usable free-form hand interaction in the wild. The process where a hand gesture is performed by the user and recognized by the machine can be decomposed into three stages. First, the user encounters an interaction task, either active or passive, and yields a hand movement intention for specific semantics. Second, the intention in the user's mind drives them to perform certain hand gestures. Third, the hand pose and motion series are observed by certain sensors in the environment. Therefore, to achieve intelligent hand behavior recognition, we should delve deeper into the modeling of these three stages. Figure \ref{fig:teaser} further outlines the detailed concepts, including sensor, behavior, and semantic modeling, along with their relationship and typical research topics. For example, in this framework, heterogeneous sensors serve as different observation or measurement channels with different sensing capability (e.g., information, resolution, and error) that contribute together to the tracking of hand behaviors. Meanwhile, hand behavior, regardless of the sensing system, has intrinsic constraints (e.g., physical and temporal), mental causes, and resolution requirements for specific applications.

Following this research framework, I previously worked on topics such as optimizing hardware form for enriched hand gesture space \cite{dualring2021liang,vahf2023li}, expanding hand input capability in spatial and temporal domains \cite{drg2023liang,dualring2021liang}, sensor fusion mechanisms \cite{vahf2023li}, behavior modeling \cite{selecting2023qin,from2023yi,9757511,isp2023liu}, and intention discovery \cite{from2023yi,9757511,dualring2021liang}.
For example, I proposed a series of novel approaches that deploy wearable sensors (e.g., IMUs, RF modules, and acoustic modules) on hands and fingers as individual channels or complements to the vision channel. These sensors, following an inside-out sensing scheme, capture the local features (e.g., the acceleration) of hand/finger segments with high sample rate and high sensitivity, and can perfectly complement vision-based solutions that provide global observation. I also work on understanding and modeling of users’ natural hand interaction behavior so as to facilitate intelligent interaction techniques that could better understand the user’s intention and allow the user to interact in a natural manner. In the following sections, I summarize my previous research topics and achievements and illustrated the future research directions. 

%% file: 2_previous_work.tex
\section{Previous Research Topics}

\subsection{Towards Optimizing Device Form for Enriched Hand Gesture Space}
An ideal interactive device is expected to be with compact sensor form while being capable to support rich hand interaction space. Although previous work investigated wearable devices such as IMU rings \cite{10.1145/3332165.3347947} and finger sleeves \cite{10.1145/3332165.3347865}, they were limited to specific applications (e.g., touch detection). To have a better representation of hand gestures, an observation is that, hand gestures are often driven by the the movement of the thumb and the index finger as representatives of two main hand segments. Inspired by this observation, we presented DualRing \cite{dualring2021liang}, a full-time wearable device consisted of two IMU rings and a high-frequency AC circuit, which achieved the sensing of on-body contact, subtle relative movement between fingers, and expressive hand gestures. With DualRing, a user can perform expressive within-hand gestures (e.g., cursor control on the fingertip, tapping different finger segments as keys, and sliding on different fingers) with perfect self-haptic feedback (unintrusively) as well as multiple hand-to-surface and hand-to-object interactions. As always-available ring-form device, DualRing was the first to outline the broad interaction space and the sub-spaces to showing its versatility for fine-grained and expressive hand gesture interactions along with applications. Evaluation results also demonstrated the computational efficiency and stronger sensing capability of DualRing compared with single-ring solution (e.g., over 25\% performance gain for gesture recognition). This work demonstrated a typical design practice focusing on the most fundamental information for supporting expressive hand gesture space. From industry perspective, Apple and Amazon have proposed their smart ring concept for ubiquitous input. As a leading implementation, DualRing featured a complete and efficient ubiquitous input solution that is capable to achieve pointing, confirmation, and text entry, which might bring novel experience to existing platforms.

Further, we went on to research how the device form (e.g., which devices and sensors to use) and the supported gesture set can be jointly optimized \cite{vahf2023li}. For an implicit hand-to-face gesture recognition task, we presented a novel cross-device sensing method that leverages heterogeneous channels (vocal, ultrasound, and IMU) of data from commodity devices (earbuds, watches, and rings). Under this framework, we conducted a simulation across all possible device and sensor channel combination to acquire the performance and saliency of different device form regarding different gesture sets, which provides guidance for the optimization of the device form under certain gesture sets. Our recognition model achieved an accuracy of 97.3\% for recognizing 3 gestures (excluding the "empty" gesture) and 91.5\% for recognizing 8 gestures, proving the high applicability. 

\subsection{Expanding Physical and Temporal Hand Input Resolution with Wearable Devices}

Considering vision-based hand tracking is limited in recognizing subtle and fast finger movements, we presented to leverage wearable devices and instruments to capture or yield significant local features to enable subtle hand interaction, which is well complementary to vision-based solutions. DualRing \cite{dualring2021liang} was the first work to achieve submillimeter unintrusive cursor control on the fingertip, which is a difficult problem in HCI due to the subtleness of finger movement. As the extension of DualRing’s subtle fingertip cursor interaction space, I further focused on users’ pixel-level 2D cursor control capability on the tiny fingertip surface to develop applicable techniques such as gesture keyboard \cite{drg2023liang} and touchpad [unpublished] on the fingertip. For example, DRG-Keyboard \cite{drg2023liang} featured a miniature QWERTY keyboard on the user’s index fingertip, allowing them to swipe the thumb on the index fingertip to perform word gesture typing. With DRG-Keyboard, a user can enter word gestures on their fingertip in a relaxed and haptic-dominated manner with the top-1 accuracy of 65.2\% and top-6 accuracy of 84.6\%. Evaluation study showed that participants could achieve an average input speed of 12.9 WPM, which is 68.3\% of their gesture typing speed on the smartphone, showing the good applicability. We also implemented a full-functional fingertip touchpad and investigated users’ control capability of fingertip cursor for simulating touchpad interaction. Results from the Fitts’ Law study and the usability study showed users were generally capable of performing accurate pointing (e.g., a 20 px × 20 px object on a 1080P display), showing the applicability of DR-Pad in real-world tasks. 

We also investigated novel approaches to enhance the sensing capability of vision-based hand tracking system with auxiliary optical construction. For example, we presented ShadowTouch \cite{10.1145/3586183.3606785}, a novel approach leveraging wrist-mounted illuminants, which cast shadows of fingers onto the near surface, to construct shadow features amplifying the subtle vertical movement of the fingers, achieving an average accuracy of 99.0\% for recognizing the touch state of independent fingers. Compared with state-of-the art hand tracking techniques (Quest Hand Tracking 2.0), ShadowTouch achieved a significantly higher resolution in differentiating subtle near-surface touch states (e.g., 2mm V.S. 9.8mm). Such an enhancement technique is well complementary to vision-based hand tracking in sensing subtle near-surface finger movement to support subtle and accurate free-form hand-to-surface touch interaction.

\subsection{Modeling and Recognizing Input Intention from Continuous Hand Motion Series}
Recognizing intentions from continuous hand input series easily suffers from Midas touch problem \cite{wu2016visual} (meaning unintended behaviors being wrongly reported as input intentions) especially when the gestures are performed in a fast and relaxed manner. We chose a typical mid-air typing scenario with dense hand input intention to investigate Midas touch problem for mid-air touch where the haptic feedback is absent \cite{from2023yi,9757511}. Traditional solutions to interact with virtual surface treated touch as detecting a collision between the finger and the interface. However, due to lack of feedback, such detection algorithms perform worse when the user is experiencing dense input (e.g., in a text entry task). In our research, we examined user's 3D typing behavior on virtual keyboards, finding that the participants perceived the projection of the lowest point during a tap as the target location and that inferring taps based on the intersection between the finger and the keyboard was not applicable. We proposed a touch prediction model based on finger kinematics, achieving a real-time detection F-1 score of 97.7\%. Based on the touch detection model, we proposed a novel input prediction algorithm that took the uncertainty in touch detection into calculation as probability, and performed probabilistic decoding that could tolerant false detection. User study showed with our technique, the user reached a single-finger typing speed of 29.6 WPM, which is 51\% faster compared with traditional text entry systems. From the results above, we found explicitly modeling of hand motion in temporal domain could achieve significantly higher accuracy (e.g., 97.7\%) in recognizing certain gestures patterns, thus significantly improving the efficiency of downstream tasks.

\subsection{Interacting Hands in Heterogeneous Environments}

The value to achieve ubiquitous intelligent hand interaction is reflected as the environments, sensing schemes, and interaction purposes are changing dynamically. In the future smart spaces, heterogeneous sensors are deployed in the environments to track human behaviors, while users can wear lightweight devices to minimize their physical burden. To this concept, IoT is a typical scenario where the user interacts their hands with surrounding objects in a free and dynamic manner. In previous work, I researched the preference and practice of hand gestures to interact with different targets in the scene, along with the sensing feasibility \cite{isp2023liu,selecting2023qin,liang2021auth,handsee2019yu}. For example, we investigated how a user embeds hand gestures into the configuration and instant control of IoT devices \cite{isp2023liu}, revealing users' preference of either binding shortcuts to certain hand gestures or using relaxed (fuzzy) hand gestures for target indication. Regarding target selection in IoT scenarios, we designed and implemented an accurate selection mechanism based on hand-held phone occlusion, achieving an average
pointing error of 1.28$^\circ$ \cite{selecting2023qin}. Besides active hand gesture input, passive states of hand also serve as a important role in conveying context information. For example, in HandSee \cite{handsee2019yu} and Auth+Track \cite{liang2021auth}, we recognized the gripping state of the hand on the smartphone as context information for UI adaptation and authentication state tracking.

%% file: 3_future_work.tex
\section{Current and Future Work}

My ongoing work focused on free-form hand-to-surface and hand-to-body interaction based on auxiliary wearables with sensor fusion and machine learning methods, aiming to turn every unmodified surface and human skin into or beyond a full-functional touch interface (as the capacitive touchscreen) as well as facilitating free-form hand gestures that are highly configurable. Specifically, the key problems I am targeted in includes: (1) Arbitrary touch on unmodified surface. Detecting arbitrary form of touch on unmodified surface, meaning to recognize the contact state of each finger as well as out-of-surface features such as touching posture and finger tilt angle, with which interactions with complex semantics such as multi-point touch and slide, different functions with different touching finger, and continuous control by tilt angle (similar to the interactions proposed by HandSee\cite{handsee2019yu}). (2) On-body interface. Robust recognition of hand-to-body event, such as pinching and hand-to-hand touching, could facilitate on-body interaction space with self-haptic feedback, which is efficient in mobile, immersive, and attention-free scenarios. Users’ interaction behavior of on-body interface is also worthy of exploring. (3) User-defined gesture space. For people with physical constraint in interaction (e.g., motor impairment or special scenarios), allowing them to define the gesture space for their unique capability that can be recognized with compact wearable sensors is beneficial in efficiency and naturalness. Understanding of how the user tend to define the gestures and finding a unified representation (e.g., a closed sensor set) in sensing those gestures is of great importance (4) Response delay and accuracy for transient hand gestures. Investigating the effect of recognition accuracy and delay for transient hand gestures such as pinching and waving hands could be beneficial to optimize the user experience and the recognition system jointly. The potential solutions to these problems fall into a form with the combination of customized cameras (e.g., one with high frame rate), wearable sensors (e.g., IMUs, RF modules, and microphones), and wearable instruments (e.g., an active IR light source and an ultrasound source) by hybridizing the sensing data from these sensors with mathematical models (e.g., error/filtering model like Kalman Filter) and machine learning methods. Such work was funded by Beijing Municipal Science and Technology Commission. In the future, I hope to explore and develop more fundamental hand interaction techniques, with which a user can interact with everything around them seamlessly, for the next generation interface. 

For a further goal, I aim to propose a unified computational model based on my research framework (Figure \ref{fig:teaser}) to represent heterogeneously distributed sensors in sensing pre-defined or user-defined free-form hand gestures. The key observation of such model is to decouple human behavioral factors and the sensing method. In the model, human behavioral data could be represented in a meta manner (e.g., keypoints or parameterized models like MANO \cite{romero2022embodied}, probably collected from high-resolution system like OptiTrack) with minimal reconstruction loss, while data streams from different sensors can be abstracted with intrinsic (e.g., sample rate, lens parameters) and extrinsic (e.g., position, orientation, and following state) parameters, with which the original data streams are capable to perform differentiable transformation. In such representation, simulated sensor data could be differentiably rendered from the meta behavioral data so that form-level optimization for sensors (e.g., finding the optimal sensor placement, property, and combination) could be performed. Currently, I am working on the theoretical basis of such computational model as well as building the differentiable rendering system, with which further investigation on unified multimodality sensing could be conducted. Moreover, I aim to model the representation and transition between the meta behavioral data and the intrinsic semantics (similar with the text-to-motion model\cite{zhang2023t2m}). Building a generative pretrained model for hand behavior would be beneficial for numbers of valuable applications such as zero-shot gesture recognition, motion synthesis, and daily activity discovery. I hope my effort could help to build an ecology to HCI where unified human-factor data and parallel semantic labels could be accumulated (e.g., from labs and different industry individuals such as film and game companies) and shared to achieve the value from the big-data perspective.


%% file: 4_conclusion.tex
\section{Conclusion}

The advancement of ubiquitous devices and scenarios has created a demand for the next generation of natural user interfaces (NUI) with human hand as the major input channel. My work aims at bridging the gap of existing vision-based hand tracking solutions to achieve highly usable hand interaction in the wild. My research is conducted under a research framework which explicitly outlines the concepts and relations of sensor, behavior, and semantic in the life cycle of an ideal hand interaction system. I believe such a research framework, along with the research topics I have conducted, would enlighten further research to facilitate ubiquitous intelligent hand interaction in the future.